\documentclass[a4paper,11pt]{article}
\usepackage{pos}

\usepackage{multicol}
\usepackage{subcaption}

\title{Determination of the Gradient Flow Scale $t_0$ from a Mixed Action with Wilson Twisted Mass Valence Quarks}
\ShortTitle{Determination of $t_0$ from a Mixed Action with Wilson Twisted Mass Valence Quarks}

\author*[a]{Alejandro Saez}
\author[a,b,c]{Alessandro Conigli}
\author[d]{Julien Frison}
\author[a]{Gregorio Herdo\'iza}
\author[a]{Carlos Pena}

\affiliation[a]{Department of Theoretical Physics, Universidad Autónoma de Madrid, 28049 Madrid, Spain \\
and Instituto de Física Teórica UAM-CSIC, c/ Nicolás Cabrera 13-15 \\
Universidad Autónoma de Madrid, 28049 Madrid, Spain}
\affiliation[b]{Helmholtz Institute Mainz, Johannes Gutenberg University, Mainz, Germany}
\affiliation[c]{GSI Helmholtz Centre for Heavy Ion Research, Darmstadt, Germany} 
\affiliation[d]{ZPPT/NIC, DESY Zeuthen, Platanenallee 6, 15738 Zeuthen, Germany}

\emailAdd{alejandro.saezg@uam.es}

\abstract{We perform the scale setting procedure of a mixed action
  setup consisting of valence Wilson twisted mass fermions at maximal
  twist on CLS ensembles with $N_f=2+1$ flavours of $O(a)$-improved
  Wilson sea quarks. We determine the gradient flow scale $t_0$ using
  pion and kaon masses and decay constants in the isospin symmetric
  limit of QCD as external physical input. We employ model variation
  techniques to explore the systematic uncertainties in the extraction
  of the ground state signal of lattice observables, as well as for
  the continuum-chiral extrapolations. We observe that the combined
  analysis of the mixed action data with that based on $O(a)$-improved
  Wilson valence quarks, provides an improved control of the
  extrapolation of $t_0$ to the physical point.  }

\FullConference{The 40th International Symposium on Lattice Field Theory (Lattice 2023)\\
July 31st - August 4th, 2023\\
Fermi National Accelerator Laboratory\\}

\begin{document}
\maketitle

\section{Introduction}

The validity of the Standard Model of particle physics has been
verified in multiple ways over the years, through the consistency
between its theoretical predictions and the corresponding experimental
measurements. However, it is expected that the Standard Model is only
an effective theory valid up to some energy scale. Searches for
physical phenomena that deviate from its predictions are one of the
main objectives in contemporary fundamental physics. One of the areas
where New Physics is expected to appear is the quark-flavor sector of
the Standard Model. A precise theoretical determination of the strong
interaction effects is needed in order to reduce the 
oretical uncertainties on flavour observables, and to uncover possible inconsistencies between theory and experiment. In this context, Lattice Field Theory is the key tool for calculating non-perturbative QCD contributions from first principles.

We consider a lattice setup \cite{mixed-action, mixed-action2, mixed-action3, mixed-action4, mixed-action5} aimed to address the leading systematic uncertainties affecting charm-quark observables. It is based on a mixed-action regularisation consisting of Wilson twisted mass valence quarks combined with CLS ensembles with $O(a)$-improved Wilson sea quarks \cite{CLS, CLS2}. Furthermore, the slowing down of the sampling of topological sectors at fine lattice spacings can be tamed by the use of open boundary conditions in the time direction \cite{open-boundary}. 

A feature of this lattice regularisation is that when valence twisted mass fermions are tuned to maximal twist, an automatic $O(a)$ improvement -- up to residual mass effects from $u,d,s$ sea quarks -- can be achieved \cite{mixed-action2, tmQCD} . This is of particular relevance when working with heavy quarks. In this work we present an update of the use of this lattice formulation in the light (up/down) and strange quark sectors. This is a necessary step before studying heavy quark physics, since a matching between the valence quark masses and the $N_f=2+1$ flavors in the sea is needed. Finally, we can carry out an independent computation of light-quark observables such as the pion and kaon decay constants and perform the scale setting using the gradient flow scale $t_0$.

\section{Sea-quark sector: $O(a)$-improved Wilson fermions}

In the sea sector, we employ the set of CLS ensembles with open boundary conditions in the time direction collected in Table \ref{tab:1}. They employ the Lüscher-Weisz gauge action with $N_f=2+1$ flavours of non-perturbatively $O(a)$-improved Wilson quarks.

These ensembles follow a chiral trajectory defined by a constant trace of the bare sea quark mass matrix
\begin{equation}\label{eqn:chiral_traj}
    \rm{Tr}\left(M_q\right)=m_u+m_d+m_s=\rm{constant}.
\end{equation}
This ensures that the improved bare coupling $\tilde{g}_0$ is kept constant up to $O(a^2)$ effects when varying the quark masses at a fixed coupling $\beta$. However, since this condition is defined for the bare quark masses, we instead opt to follow a renormalised chiral trajectory by imposing that 
\begin{equation}\label{eqn:phi}
\phi_4=8t_0\left(m_K^2+\frac{1}{2}m_{\pi}^2\right)=8t_0m_K^2+\frac{1}{2}\phi_2,
\end{equation}
is kept fixed to its physical value. This choice corresponds at LO in $\chi$PT to fixing the sum of the renormalized quark masses, since
\begin{equation}
    \phi_4\propto {\rm m_{u}^{R}+m_{d}^{R}+m_{s}^{R}}.
\end{equation}
The target chiral trajectory can be reached through a mass-shift of the simulated quark masses by Taylor expanding an observable $O$ in the bare quark masses \cite{BKS} as follows
\begin{equation}\label{eqn:mass-shift}
    \left<O(m'_u,m'_d,m'_s)\right>=\left<O(m_u,m_d,m_s)\right>+\sum_q(m'_q-m_q)\frac{d\left<O\right>}{dm_q}.
\end{equation}
Following \cite{thesis}, the sum in eq.~(\ref{eqn:mass-shift}) is restricted to $q=s$ since shifting only the strange quark mass leads to an improved precision in the target observables.
An educated guess for the physical value $t_0^{\rm ph}$,
\begin{equation}\label{eqn:t0*}
  \sqrt{t_0^{\rm iter}}=0.1445(6)\;\text{fm}.
\end{equation}
is selected as input of the final iteration leading to the value of $\phi_4$ to which all observables are then mass-shifted. 
This specific value in  eq.~(\ref{eqn:t0*}), is the outcome of the first steps of an {\it
    iterative} procedure starting from an initial guess of $t_0$,
  which is chosen free of uncertainties, and iterates the complete scale
  setting analysis, including correlations, until convergence to the value of $t_0$ is reached. The steps of this procedure are illustrated in the right
  panel of Fig. \ref{fig:t0_compar}.
  Using the isospin symmetric (isoQCD) values of the  pion and kaon meson masses recommended in Ref.~\cite{FLAG21},
\begin{align}\label{eqn:mass_input}
    m_{\pi}^{\rm isoQCD}&=m_{\pi^0}^{\rm exp}=134.9768(5)\;\text{MeV},\\
    m_K^{\rm isoQCD}&=m_{K^0}^{\rm exp}=497.611(13)\;\text{MeV},
\end{align}
leads to the value for $\phi_4^{*}=1.101(9)$. 

\begin{table}
\begin{center}
\begin{tabular}{c c c c c} 
\hline
$\beta$ & $a$ [fm] & id & $m_{\pi}\;[\text{MeV}]$ & $m_K\;[\text{MeV}]$ \\
\hline
3.40 & 0.086 & H101 & 420 & 420 \\
 & & H102 & 350 & 440 \\
 & & H105 & 280 & 460 \\
3.46 & 0.076 & H400 & 420 & 420 \\
3.55 & 0.064 & N202 & 420 & 420 \\
 & & N203 & 340 & 440 \\
 & & N200 & 280 & 460 \\
 & & D200 & 200 & 480 \\
3.70 & 0.050 & N300 & 420 & 420 \\
 & & N302 & 340 & 440 \\
 & & J303 & 260 & 470 \\
\hline
\end{tabular}
\caption{\label{tab:1}$N_f=2+1$ CLS ensembles \cite{CLS} used in the sea sector. These ensembles employ non-perturbatively $O(a)$-improved Wilson fermions and open boundary conditions in the time direction.}
\end{center}
\end{table}

\section{Valence-quark sector: Wilson twisted mass fermions}

In the valence sector we use the Wilson twisted mass (Wtm) fermion action at maximal twist. This regularisation adds a chirally rotated mass term to the massless Wilson Dirac operator $D_{\rm W}$ including the Sheikholeslami-Wohlert term,
\begin{equation}\label{eqn:Wtm}
  %D_{W}+m_q^{\rm val}\rightarrow
  D_{\rm Wtm}=D_{\rm W}+m_0^{\rm val}+i\gamma_5\mu_q^{\rm val}\;,
\end{equation}
where the $m_0^{\rm val}$ mass enters in the  standard subtracted quark
mass, $m_q^{\rm val}=m_0^{\rm val}-m_{\rm cr}$, and $\mu_q^{\rm val}$ is the twisted mass parameter.
Physical observables constructed from the Wtm Dirac operator are $O(a)$-improved -- save for residual lattice artefacts of $O(a g_0^4 \rm{Tr}\left(M_q\right))$ arising from sea quark masses -- once the maximal twist condition is fixed by tuning the hopping parameter, $\kappa^{\rm val}=(2am_0^{\rm val}+8)^{-1}$,  so that the light ($u,d$) valence PCAC quark mass vanishes, $m_{12}^{\rm val}=0$, on each ensemble. 

\section{Matching sea and valence sectors}

\begin{figure}
\begin{center}
  \includegraphics[width=.5\linewidth]{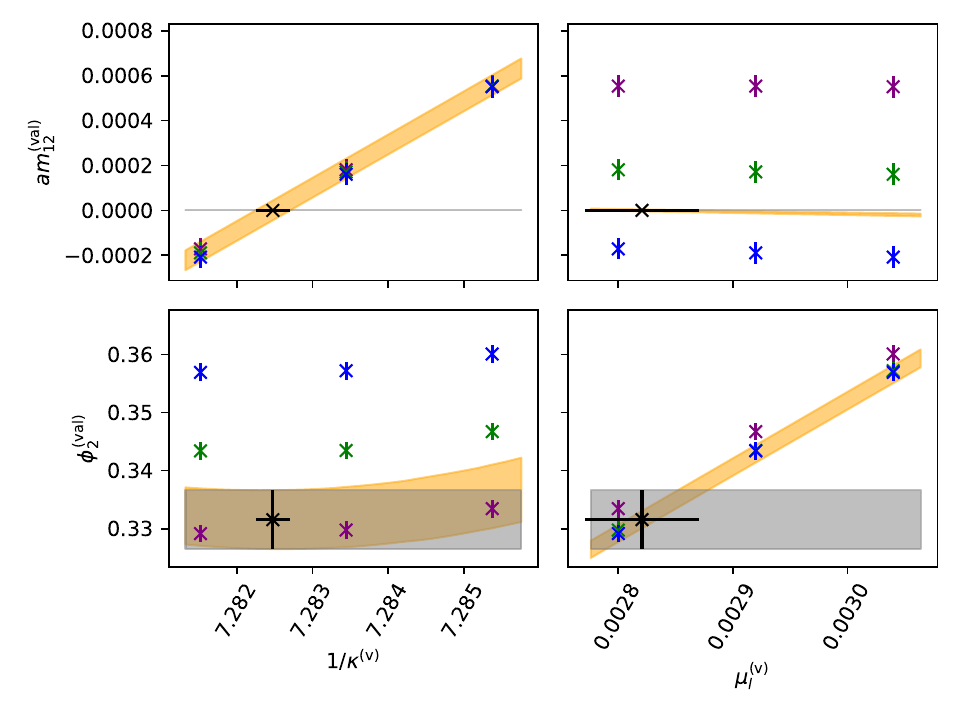}
\end{center}
\caption{Top row: Light ($u,d$) valence PCAC quark mass from the valence {\it grid} of points -- in the hyperplane of input parameters $(\kappa^{\rm val}, a\mu_l^{\rm val}, a\mu_s^{\rm val})$ -- interpolated to the maximal twist condition, $m_{12}^{\rm val}=0$ . Bottom row: $\phi_2^{\rm val}$ along the valence grid, interpolated to the sea value $\phi_2^{\rm sea}$. The sea sector parameters correspond to those of Table \ref{tab:1} for ensemble H105. The orange band in both figures represents the interpolation along the grid of valence parameters, while the horizontal grey line and band represent the target value to which we want to interpolate both observables. In the case of $am_{12}^{\rm val}$, it is set to zero, and for $\phi_2^{\rm val}$ to $\phi_2^{\rm sea}$. The interpolation of $\phi_4^{\rm val}$ is carried out in a similar way to that of $\phi_2^{\rm val}$.}
\label{fig:matching_m12}
\end{figure}

Since we are using a mixed action setup, to recover the unitarity of the theory in the continuum limit, we require the matching of the sea and valence quark masses. This matching can also be imposed in terms of the observables $\phi_2,\phi_4$ defined in eq. (\ref{eqn:phi}), depending on the kaon and pion masses, 
\begin{equation}\label{eqn:match}
\phi_2^{\rm val}\equiv\phi_2^{\rm sea}\,,\quad \phi_4^{\rm val}\equiv\phi_4^{\rm sea}.
\end{equation}
More specifically, we employ a set of measurements in the valence hyperplane $(\kappa^{\rm val}, a\mu_l^{\rm val}, a\mu_s^{\rm val})$ that allows to perform small interpolations to the valence parameters that satisfy eq. (\ref{eqn:match}), in addition to simultaneously imposing the maximal twist condition, $m_{12}^{\rm val}=0$. The interpolation fit functions can be based on LO $\chi$PT. The matching procedure is illustrated in Fig. \ref{fig:matching_m12}.

Having determined the valence {\it matching} parameters $(\kappa^{\rm val,*},\;a\mu_l^{\rm val,*},\;a\mu_s^{\rm val,*})$ at which the maximal twist and the matching conditions in eq. (\ref{eqn:match}) are simultaneously satisfied, it is then possible to also interpolate the valence pion and kaon decay constants to the same {\it matching} point.

We note that, throughout the analysis, finite volume effects are corrected using LO $\chi$PT \cite{CDH}. These corrections are found to be smaller than the size of the statistical uncertainties for all observables and ensembles. The extraction of the ground state signal of lattice observables is based on a model variation over the Euclidean time fit intervals.

\section{Scale setting and determination of $t_0^{\rm ph}$}

The scale setting is carried out in terms of a linear combination of the pion and kaon decay constants in units of $t_0$~\cite{BKS},
\begin{equation}\label{eqn:t0fpik}
\sqrt{8t_0}f_{\pi K}=\sqrt{8t_0}\times\frac{2}{3}\left(f_K+\frac{1}{2}f_{\pi}\right).
\end{equation}
Up to logarithmic corrections, this quantity remains constant in SU(3) $\chi$PT at NLO along the renormalized chiral trajectory defined by $\phi_4\equiv\phi_4^*$.

We will consider three sets of data in the scale setting analysis: (i) the {\it unitary} setup where the same $O(a)$-improved Wilson Dirac operator is used in the sea and valence sectors, which are referred to as the ``Wilson'' case; (ii) the mixed action setup after the matching procedure, referred to as ``Wtm''; and (iii) the {\it Combined} set of data, in which the two previous sets are analysed simultaneously by imposing the universality of continuum-limit result. More specifically, the ``Wilson'' and ``Wtm'' data sets can be analysed independently, leading to two determinations of the physical value of $t_0^{\rm ph}$. For the {\it Combined} case, a simultaneously fit of both data sets with independent sets of parameters characterising cutoff effects is performed.
The lattice data can be parameterised as follows
\begin{equation}\label{eqn:fit-individual}
    \left(\sqrt{8t_0}f_{\pi K}\right)^{\text{latt}}=\left(\sqrt{8t_0}f_{\pi K}\right)^{\text{cont}}+\textit{c}(a,\phi_2),\;
\end{equation}
with $\textit{c}(a,\phi_2)$ a function describing cutoff effects. Several possible choices for the continuum behaviour and the cutoff effects are explored.
For the continuum mass-dependence we consider SU(3) $\chi$PT expressions at NLO
\begin{equation}\label{eqn:fit-proceedings}
    \left(\sqrt{8t_0}f_{\pi K}\right)^{\text{cont}}=\frac{p_1}{8\pi\sqrt{2}}\left[1-\frac{7}{6}L\left(\frac{\phi_2}{p_1^2}\right)-\frac{4}{3}L\left(\frac{\phi_4-\phi_2/2}{p_1^2}\right)-\frac{1}{2}L\left(\frac{4\phi_4/3-\phi_2}{p_1^2}\right)+p_2\phi_4\right]\;,
\end{equation}
where, $L(x)=x\log(x)$\;.
An alternative is to employ a Taylor expansion around the symmetric point $\phi_2^\text{sym}$,
\begin{equation}\label{eqn:fit-taylor-2}
    \left(\sqrt{8t_0}f_{\pi K}\right)^{\text{cont}}=p_1+p_2(\phi_2-\phi_2^\text{sym})^2.
\end{equation}

\begin{figure}
\centering
  \includegraphics[width=1.0\linewidth]{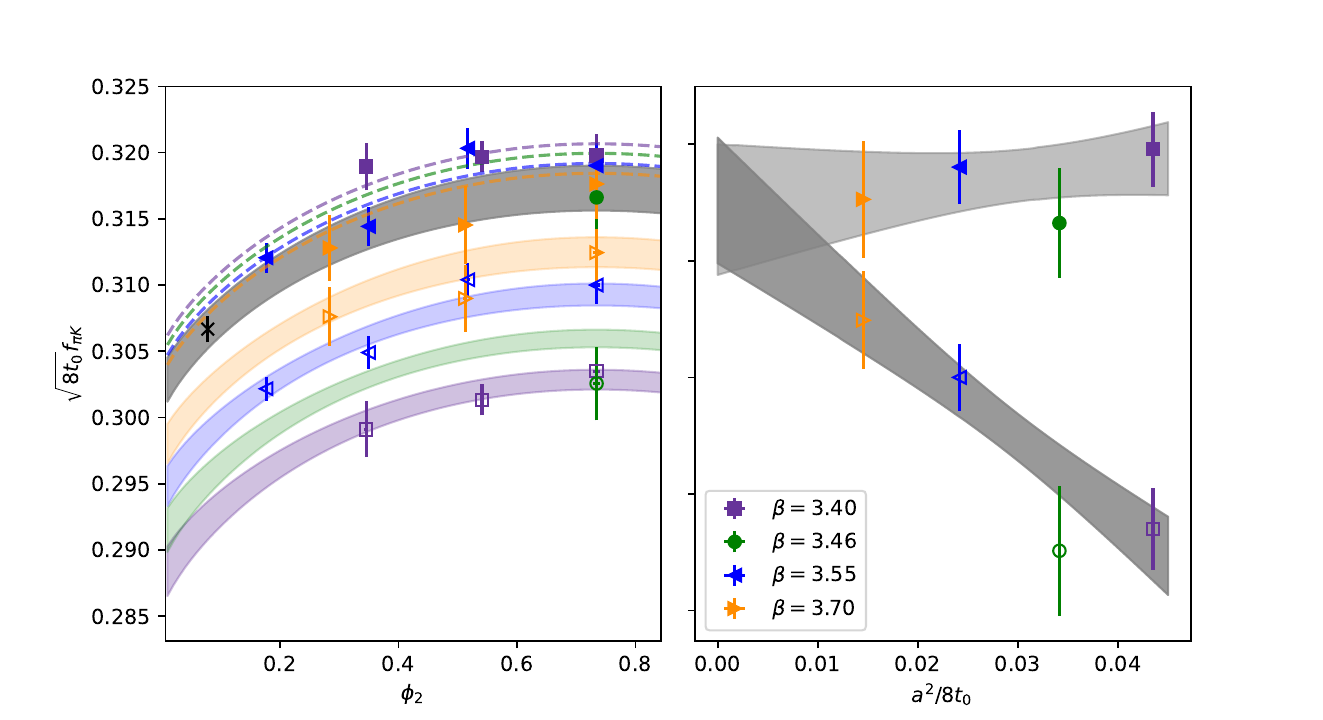}
  \caption{Left: Chiral and continuum extrapolations of $\sqrt{8t_0}f_{\pi K}$. We show the measurements for Wilson (empty points) and Wtm (filled points). The fit form in eq. (\ref{eqn:fit-proceedings}) is used for mass dependence together with eq. (\ref{eqn:cutoff}) with $c_2=c_3=0$ to parameterise the lattice spacing dependence. No cuts are applied in this specific fit. Points with the same colour refer to a common value of the lattice spacing. The grey band represents the continuum limit dependence for the {\it Combined} data set analysis, while the coloured bands represent the chiral fits at each lattice spacing for the Wilson data, and, similarly, the dashed lines  (without showing uncertainty in the fits) correspond to the Wtm case. Right: Continuum limit of symmetric point ensembles, $m_{\pi}=m_K$, with
    $\phi_2=0.740(9)$. A common continuum limit is not imposed in this fit.}
\label{fig:continuum_chiral_limit}
\end{figure}

To characterise the lattice spacing dependence, we consider
\begin{equation}\label{eqn:cutoff}
c(a,\phi_2)=c_1\frac{a^2}{t_0}+c_2\phi_2\frac{a^2}{t_0}+c_3\alpha_s^{\hat{\Gamma}}\frac{a^2}{t_0},
\end{equation}
and we examine the effect of switching on/off the different $c_i$'s. An exploratory study of the impact of including logarithmic corrections of $O(a^2\alpha_s^{\hat{\Gamma}})$
considering the smallest value of the anomalous dimensions $\hat{\Gamma}$ reported in \cite{Husung} is also incorporated in the analysis by including the $c_3$ term while setting $c_1=0$.
We observe that the quality of the fit with cutoff effects of
$O(\phi_2a^2)$ is poor for the current set of Wilson data, while a good description can be obtained  for the Wtm data. 

In addition to a variation of the functional forms, we explore the possibility of performing cuts in the data, by removing the coarsest lattice spacing $\beta=3.40$, or by cutting the  symmetric point ensembles with $m_{\pi}=420$ MeV. 

The correlations present in the Monte Carlo data are retained
throughout the analysis. In the chiral-continuum fit, the $\chi^2$
function includes the correlations among the $\sqrt{t_0}f_{\pi K}$
data points while the residual cross-correlations between
$\sqrt{t_0}f_{\pi K}$ and $\phi_2$ are neglected in the fit. We observe, however, that this leads to a tiny deviation of the expectation value of the chi-squared \cite{chiexp}, $\left<\chi^2\right>$, away from the number of degree-of-freedom, i.e. the expected value when considering a correlated fit.

\begin{figure}
\centering
  \includegraphics[width=.9\linewidth]{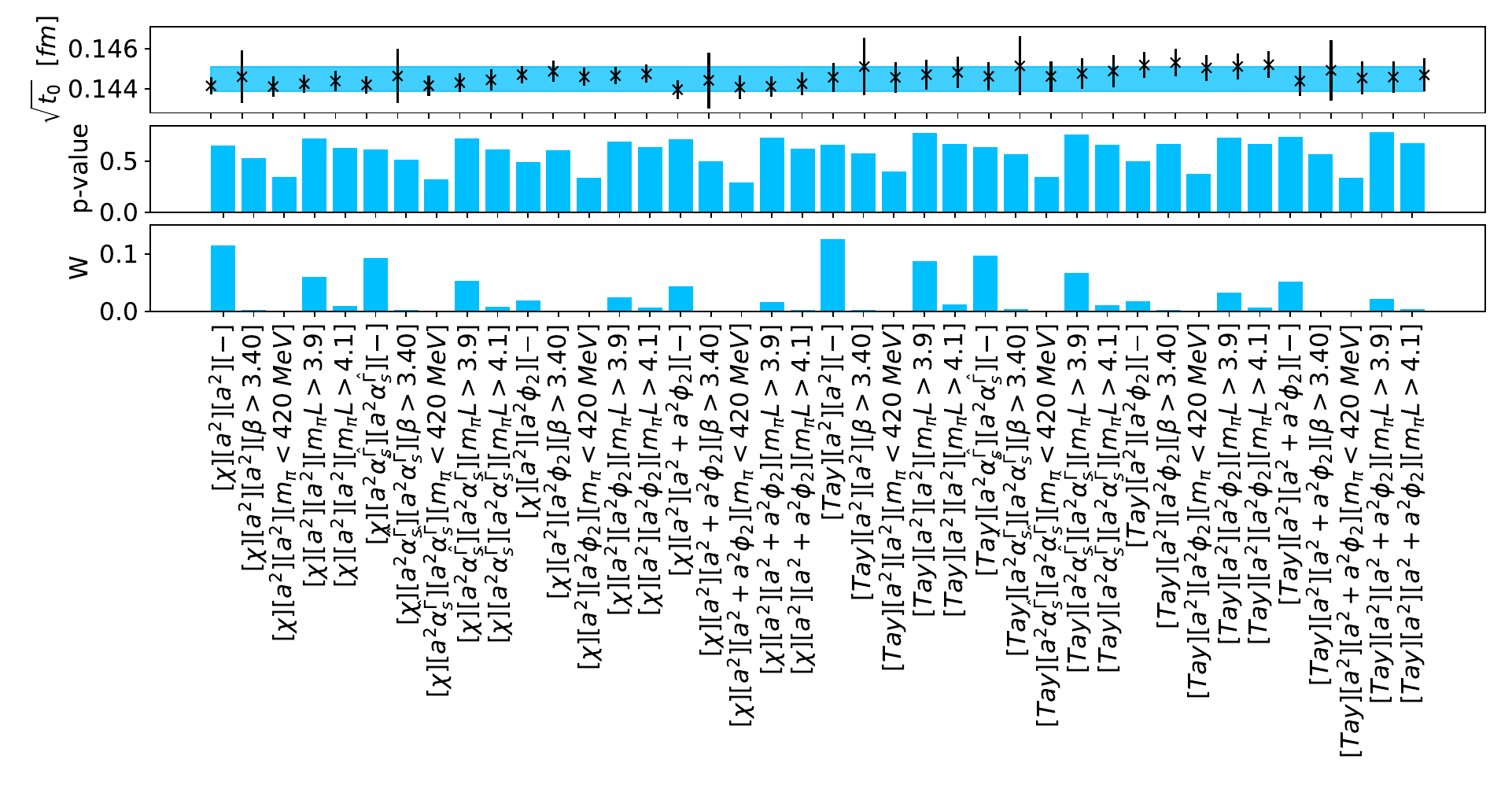}
  \caption{Results for $\sqrt{t_0^{\rm ph}}$ for the {\it Combined}
    data set and for each model considered. The p-value and $\left<
    \chi^2 \right>$ are computed following \cite{chiexp}. The blue
    band represents the model average following \cite{ethan_neil}
    where the systematic error contribution from the model variation
    has been included. The horizontal axis refers to the considered
    models: the first tag ([$\chi$] or [Tay]) refers to the continuum
    parameterisation according to eq. (\ref{eqn:fit-proceedings}) or
    eq. (\ref{eqn:fit-taylor-2}), respectively; the second and third
    tags label the cutoff effects of the Wilson and Wtm subsets of
    data, respectively ([$a^2$] if only $c_1$ in
    eq. (\ref{eqn:cutoff}) is switched on,
    [$a^2\alpha_s^{\hat{\Gamma}}$] if only $c_3$ is used,
    [$\phi_2a^2$] if $c_2$ is used, or [$a^2+\phi_2a^2$] if both $c_1$ and $c_2$ are considered); and the fourth tag corresponds to the cuts performed in both Wilson and Wtm data for the fit ([$\beta>3.40$] if only ensembles with $\beta>3.40$ are kept or [$m_{\pi}<420$\,MeV] if symmetric point ensembles are discarded). Although only a subset of the models do have a significant weight, we observe that the error band from the model averaging covers the results of all the individual models. The first model in the $x$-axis corresponds to the one illustrated in Fig. \ref{fig:continuum_chiral_limit}.}
\label{fig:t0_ph_syst}
\end{figure}

To study the systematic effects associated with model variation in the chiral and continuum extrapolations, we use the model averaging method introduced in \citep{ethan_neil} with the information criterion proposed in \cite{TIC} to take into account fits that are not fully correlated. According to this information criterion, each fit model is assigned a weight
\begin{equation}\label{eqn:weight}
W\propto\exp\left(-\frac{1}{2}\left(\chi^2-\left<\chi^2\right>\right)\right),
\end{equation} 
which allows to compute a weighted average for $\sqrt{t_0}f_{\pi K}$
over the explored models. The method also assigns a systematic
uncertainty coming from the model variation.\,\footnote{We have also
  considered this model averaging technique to extract the ground
  state signal of the relevant lattice observables  by scanning over
  Euclidean-time fit intervals.} The model variation procedure and the
corresponding model average are illustrated in
Fig. \ref{fig:t0_ph_syst}. The specific model based on eq. (\ref{eqn:fit-proceedings}) for the continuum behaviour and $c_2=c_3=0$ term in eq. (\ref{eqn:cutoff}) is shown in Fig. \ref{fig:continuum_chiral_limit}.

\begin{figure}
\centering
    \includegraphics[width=.5\linewidth]{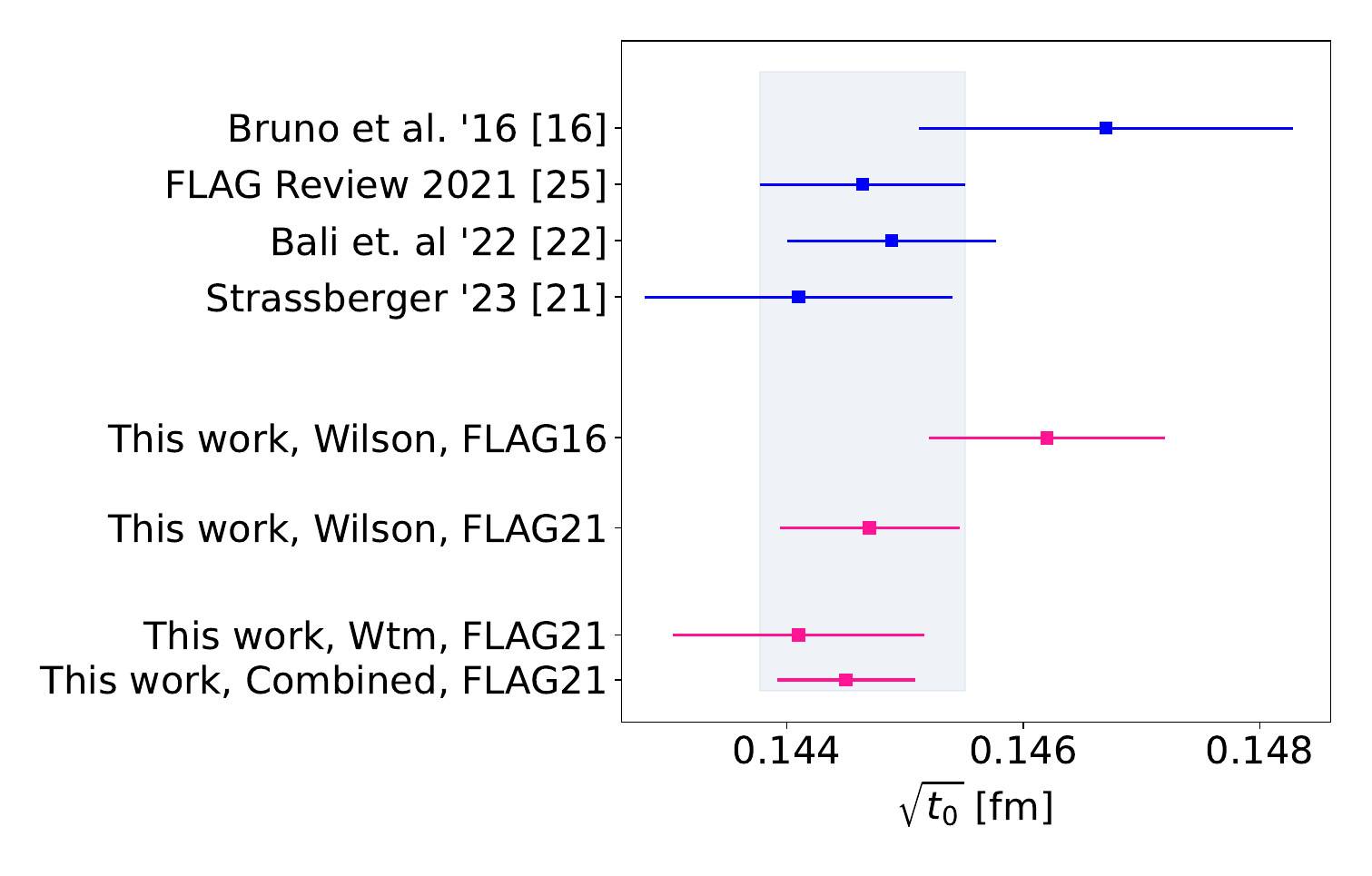}
    \includegraphics[width=.425\linewidth]{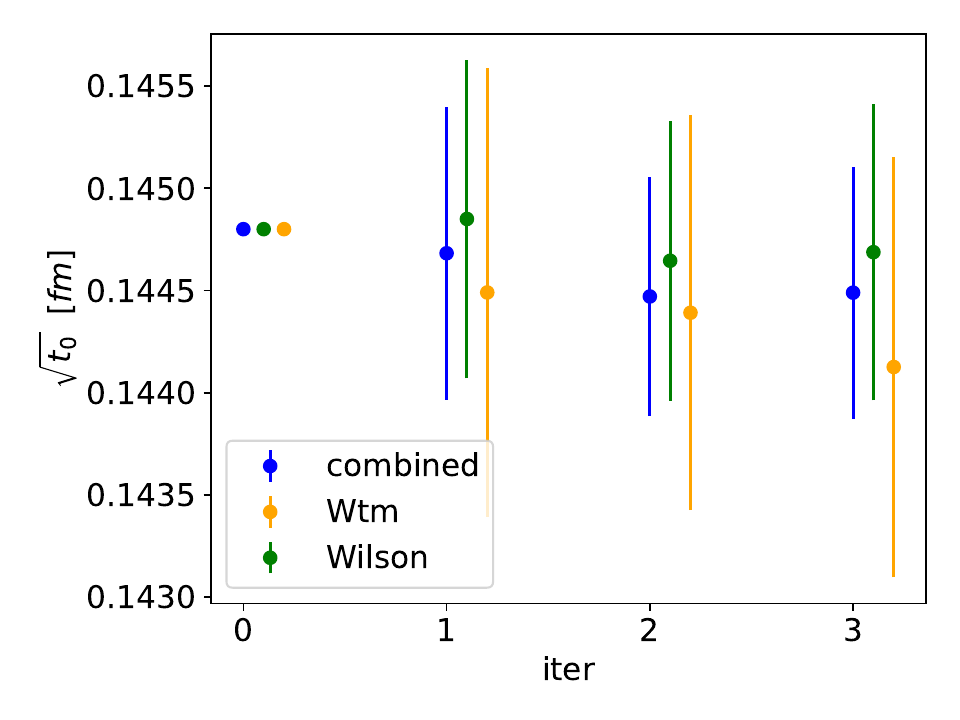}
    \caption{Left: Results for the gradient flow scale $\sqrt{t_0}$ in
      physical units based on the model average (pink points) for our
      three types of analysis (Wilson, Wtm and Combined), compared
      with the $N_f=2+1$ FLAG average \cite{FLAG21} and with other
      determinations also based on $N_f=2+1$ CLS ensembles (blue
      points). Since in Bruno et al. \cite{BKS} the FLAG16
      \cite{FLAG16} prescription was used to set in the physical
      input, we also show the result of our analysis using the Wilson
      regularisation with FLAG16 physical input ("This work, Wilson,
      FLAG16") instead of FLAG21 \cite{FLAG21} ("This work, Wilson,
      FLAG21"). The CLS results, Bali et al. \cite{RQCD:2022xux} and
      Strassberger \cite{thesis}, use FLAG21 \cite{FLAG21} physical
      input. Right: results of $\sqrt{t_0}$  for the three types of
      analysis at each iteration of the scale setting analysis. The
      initial iteration starts from a $t_0$ value without error and
      the $i$-th iteration starts with the value of $t_0^{\rm ph}$ of
      the $(i-1)$-th iteration. After four iterations of the scale
      setting analysis we observe signs of convergence for the value of $t_0$.}
\label{fig:t0_compar}
\end{figure}

Once $\sqrt{8t_0}f_{\pi K}$ is determined at the physical point (i.e. continuum limit result with physical pion and kaon masses), using a prescription for the isoQCD values of the pion and kaon decay constants~\cite{FLAG21} 
\begin{equation}\label{eqn:f_inputs}
    f_{\pi}^{\rm isoQCD}=130.56(13)\;\text{MeV},\;\;f_K^{\rm isoQCD}=157.2(5)\;\text{MeV},
\end{equation}
%5
as physical inputs, it is possible to determine the physical value of the gradient flow scale $t_0^{\rm ph}$. Our results for the three types of analysis (Wilson, Wtm and Combined), together with a comparison with other studies also based on $N_f=2+1$ CLS ensembles, are presented in Fig. \ref{fig:t0_compar}. We observe a shift of the central value of $t_0$ with respect to the result of Bruno et al. \cite{BKS} --- which is based on a similar set of ensembles than the one employed in this work --- depending on whether the physical inputs of FLAG21 \cite{FLAG21} or FLAG16 \cite{FLAG16} are considered.

\section{Conclusions}

We have presented an update of a scale setting procedure based on physical inputs for the pion and kaon decay constants, using a mixed action consisting of twisted mass valence quarks on CLS $O(a)$-improved sea quarks \cite{lattice22}. We have explored the systematic effects associated to model variation in the chiral and continuum limits, and demonstrated the effectiveness of combining the Wilson and mixed action (Wtm) calculations to improve the precision of $t_0$.
In the near future we plan to extend this study by adding ensembles at finer lattice spacings and with physical pion masses, which will allow to consider an analysis where physical input from only the pion decay constant is employed in the determination of $t_0$. Furthermore, we plan to determine the light quark masses and to pursue our charm physics project based on this mixed action approach \cite{charm_paper, Alessandro, Julien}.

\section*{Acknowledgements}

\noindent We are grateful to our colleagues in the Coordinated Lattice
Simulations (CLS) initiative for the generation of the gauge field
configuration ensembles employed in this study. We acknowledge PRACE
for awarding us access to MareNostrum at Barcelona Supercomputing
Center (BSC), Spain and to HAWK at GCS@HLRS, Germany. The authors
thankfully acknowledge the computer resources at MareNostrum and the
technical support provided by Barcelona Supercomputing Center
(FI-2020-3-0026). We thank CESGA for granting access to Finis Terrae
II. This work is partially supported by grants PGC2018-094857-B-I00
and PID2021-127526NB-I00, funded by MCIN/AEI/10.13039/501100011033 and
by “ERDF A way of making Europe”, and by the Spanish Research Agency
(Agencia Estatal de Investigaci\'on) through grants IFT Centro de
Excelencia Severo Ochoa SEV-2016-0597 and No CEX2020-001007-S, funded
by MCIN/AEI/10.13039/501100011033. We also acknowledge support from
the project H2020-MSCAITN-2018-813942 (EuroPLEx), under grant
agreement No. 813942, and the EU Horizon 2020 research and innovation programme, STRONG-2020 project, under grant agreement No 824093.

\end{document}